\RequirePackage{fix-cm}

\documentclass[twocolumn,epjc3]{svjour3}
\usepackage{amsmath}
\usepackage[english]{babel}
\usepackage{graphicx}
\usepackage{psfrag}
\usepackage{cite}

\def\as{\alpha_{s}}

\def\asb{\bar{\alpha}_{s}}
\def\Ltau{L}

\def\asl{\alpha_{s}\Ltau}

\def\d{\hbox{d}}

\smartqed  
\RequirePackage{graphicx}
\journalname{Eur. Phys. J. C}
\begin{document}

\title{Power corrections in the dispersive model 
for a determination of the strong coupling constant from the thrust distribution}

\author{Thomas~Gehrmann\thanksref{e1,addr1}, Gionata~Luisoni\thanksref{e2,addr2,addr3}
        \and
        Pier~Francesco~Monni\thanksref{e3,addr1} 
}

\thankstext{e1}{e-mail: thomas.gehrmann@uzh.ch}
\thankstext{e2}{e-mail: luisonig@mpp.mpg.de}
\thankstext{e3}{e-mail: pfmonni@physik.uzh.ch}

\institute{Institut f\"ur Theoretische Physik, Universit\"at Z\"urich,
Winterthurerstrasse 190,\\CH-8057 Z\"urich, Switzerland\label{addr1}
           \and
Institute for Particle Physics Phenomenology, University of Durham,\\
Science Laboratories, South Rd, Durham DH1 3LE, UK \label{addr2}
	   \and
Max-Planck-Institut f\"ur Physik, F\"ohringer Ring 6,\\D-80805 M\"unchen, Germany \label{addr3}}

\date{Received: date / Accepted: date}

\maketitle

\begin{abstract}
In the context of the dispersive model for non-perturbative corrections, we extend the leading renormalon subtraction to NNLO for the thrust distribution in $e^+e^-$ annihilation. 
Within this framework, using a NNLL+NNLO perturbative description and including bottom quark mass effects to NLO, we 
analyse data in the centre-of-mass energy range $\sqrt{s}=14-206$~GeV in view of a simultaneous determination of the strong coupling constant and the non-perturbative parameter $\alpha_0$. 
The fits are performed by matching the resummed and fixed-order predictions both in the R and the log-R matching schemes. The final values in the R scheme are $\alpha_s(M_Z) = 0.1131^{+0.0028}_{-0.0022}$, $\alpha_0(2~{\rm GeV}) = 0.538^{+0.102}_{-0.047}$.

\end{abstract}
\section{Introduction}
\label{intro}
The thrust ($T$) distribution~\cite{thrust} in $e^+e^-$ annihilation is one of the most precisely measured event shape observable at electron-positron
colliders~\cite{aleph,opal,l3,delphi,sld,jade,tasso}. Owing to its infrared and collinear safety it can be computed systematically in perturbation theory. It spans the range between the kinematical
configuration of two collimated back-to-back jets (dijet limit) and a perfectly spherical final state. The deviation from a dijet configuration is proportional to the strong coupling constant
$\alpha_s$ and thus allows for a  precise determination of its value at a given scale.
Fixed-order QCD predictions for this observable, as well as for other three-jet event shape distributions, are known
to next-to-next-to-leading order (NNLO)
accuracy~\cite{GehrmannDeRidder:2007jk,Weinzierl:2009nz}. These results were used in~\cite{Dissertori:2007xa,jaquier,Maxwell:2011dy} to perform
precise fits of the strong coupling constant. Further fits based on the analysis of jet rates have been performed in~\cite{asjet,Schieck:2012mp,Frederix:2010ne}. 
Fixed-order calculations give a reliable description of the experimental data when three or more final state jets are created. On the other hand, they diverge in the dijet region (where $T\rightarrow
1$) because of large logarithms of the form $\alpha_s^n \ln^{2 n} (1-T)$. In order to obtain a reliable theory prediction for the cross section, these logarithmic terms have to be summed to
all orders in the strong coupling. The resummed cross section is then matched to the fixed-order result so as to obtain a prediction holding all over the kinematical range spanned by the thrust. The
matching procedure is to be defined taking care of avoiding double counting with the fixed-order cross section~\cite{CTTW,Jones:2003yv}. 
Resummed calculations to next-to-leading logarithmic (NLL) accuracy are available for a wide class of event shape observables~\cite{CTTW,broadenings,Becher:2011pf,y3}, and
various determinations of $\alpha_s$ relying on these predictions matched to NNLO
({\it i.e.} NLL+NNLO) were obtained in~\cite{Gehrmann:2008kh,davisonwebber,jadeas,Dissertori:2009ik,opalas}. The inclusion of the resummation led to an important reduction in the scale uncertainty and showed the need to go
beyond the next-to-leading logarithmic accuracy.
For the thrust cross section, the resummation of large logarithms beyond NLL has been performed both following effective theory methods~\cite{Becher:2008cf} and using a traditional
approach~\cite{Monni:2011gb}. A resummed prediction to this accuracy is also available for the heavy-jet mass~\cite{Chien:2010kc} and the jet-broadenings~\cite{Becher:2011pf,Becher:2012qc}.
Moreover, precise fits of the strong coupling have been performed using the effective theory results~\cite{Abbate:2010xh,Abbate:2012jh}.

In this paper we present a determination of $\alpha_s$ based on theoretical NNLL predictions matched to NNLO for the thrust distribution. To describe hadronisation effects, we 
extend the dispersive model~\cite{Dokshitzer:1995qm,Dokshitzer:1997ew} 
to match NNLL+NNLO and perform a subtraction of the 
leading renormalon at NNLO. 
 Bottom-quark mass effects are considered to NLO. The
theoretical parton level results are reviewed in Section~\ref{sec:Thrust distribution in perturbation theory}. In Section~\ref{sec:bottom-mass corrections} we describe the inclusion of bottom-quark
mass corrections, whereas the non-perturbative corrections and the extension of the dispersive model to match the perturbative NNLL+NNLO is described in Section~\ref{sec:Non-perturbative corrections}.
Section~\ref{sec:Determination of alpha} is dedicated to the determination of $\alpha_s$, the fit procedure and the results. Finally, Sections~\ref{sec:Comparison} and~\ref{sec:Conclusions} contain the comparison to other recent determinations of $\alpha_s$ and our conclusions.

\section{Thrust distribution in perturbation theory}
\label{sec:Thrust distribution in perturbation theory}
\label{sec:framework}
The thrust variable for a hadronic final state in $e^+e^-$ annihilation is
defined as~\cite{thrust}
\begin{align}
T=\max_{\vec{n}}
\left(\frac{\sum_i |\vec{p_i}\cdot \vec{n}|}{\sum_i |\vec{p_i}|}\right)\,,
\label{thrust}
\end{align}
where $\vec{p}_i$ denotes the three-momentum of particle $i$, with the sum running
over all particles. The unit vector $\vec{n}$ is varied to find  the
thrust direction $\vec{n}_T$ which maximises the expression in parentheses
on the right hand side. In the present paper we will mostly work with the quantity $\tau\equiv 1-T$.

\subsection{\bf{Fixed-order and resummed calculations}}
The differential thrust distribution in perturbation theory is known at NNLO~\cite{GehrmannDeRidder:2007jk,Weinzierl:2009nz}. At a centre-of-mass energy $Q$ and for
a renormalisation scale $\mu$ it reads

\begin{align}
\frac{1}{\sigma}\, \frac{\d\sigma}{\d \tau}(\tau,Q) &= \bar\alpha_s (\mu) \frac{\d A}{\d \tau}(\tau)+ \bar\alpha_s^2 (\mu) \frac{\d B}{\d \tau} (\tau,x_\mu) \nonumber\\
\label{eq:fixedordercs0}
&+ \bar\alpha_s^3 (\mu) \frac{\d C}{\d
\tau}(\tau,x_\mu) +
{\cal O}(\bar\alpha_s^4)\;,
\end{align}
where we defined
\begin{align}
\asb = \frac{\alpha_s}{2\pi}\;, \qquad x_\mu = \frac{\mu}{Q}\;,
\end{align}
and where $\sigma$ is the total perturbative hadronic cross-section for $e^{+}e^{-}\rightarrow$ hadrons.
The perturbative strong coupling constant is defined in the $\overline{\rm MS}$ scheme unless stated otherwise. In Section~\ref{sec:Non-perturbative corrections} below 
we will use a different scheme for $\alpha_s$. 

\noindent In theoretical computations it is customary to normalise the distributions to the Born cross
section $\sigma_{0}$ since, for massless quarks, the normalisation cancels all electroweak coupling factors. However, the experimental measurement is 
of course normalised to the total hadronic cross section $\sigma$.

The normalised thrust cross-section is defined as
\begin{align}\label{eq:Rfixed}
R_{T}(\tau)\,\equiv\,\frac{1}{\sigma}\int_{0}^{1}\frac{d\sigma\left(\tau',Q\right)}{d\tau'}\Theta(\tau-\tau^{\prime})d\tau',
\end{align}
whose fixed-order expansion has the general form
\begin{align}\label{eq:Rfixedorder}
R_{T}\left(\tau\right)\,&=\,1+R_{1}\left(\tau\right)\asb(\mu^{2})\,+\,R_{2}\left(\tau,\mu^{2}\right)\asb^{2}(\mu^{2})\,\nonumber\\
&+\,R_{3}
\left(\tau, \mu^ {2} \right)\asb^{3}(\mu^{2}).
\end{align}
The fixed-order coefficients $R_{1}$, $R_{2}$, $R_{3}$ can be obtained by integrating the distribution~(\ref{eq:fixedordercs0}) and imposing the condition
 $R_{T}(\tau_{{\rm max}},Q)=1$ to all
orders, where $\tau_{{\rm max}}$ is the maximal kinematically allowed value ($\tau_{\rm max} \simeq 0.4275$ at ${\cal O}(\alpha_s^3)$).

In the two-jet region the fixed-order distribution is enhanced by large infrared logarithms which spoil the convergence of the perturbative series. The convergence can be restored by resumming
the logarithms to all orders in the coupling constant.  The matched cross section can in general be written as
\begin{align}\label{eq:Rres}
R_{T}(\tau) = C(\alpha_{s}) \Sigma(\tau,\alpha_{s})+D(\tau,\alpha_{s}),
\end{align}
where
\begin{align}
\label{eq:csigma1}
C(\alpha_{s}) &= 1+\sum_{k=1}^{\infty}C_{k}\asb^{k},\\
\ln\Sigma(\tau,\alpha_{s}) &= \sum_{n=1}^{\infty}\sum_{m=1}^{n+1}G_{nm}\bar{\alpha_{s}}^{n}L^{m}\nonumber\\
\label{eq:csigma2}
=&\Ltau g_{1}(\asl)+g_{2}(\asl)+\frac{\alpha_s}{\pi}\beta_0g_3(\asl)+\ldots
\end{align}

\noindent where $L\equiv\ln(1/\tau)$. The function $g_{1}$ encodes all the leading logarithms, the function $g_{2}$ resums all next-to-leading logarithms and so on.
The constant terms $C_{i}$ are required to achieve a full N$^{1+i}$LL accuracy.
$D(\tau,\alpha_{s})$ is a remainder function that vanishes order-by-order in perturbation theory
in the dijet limit $\tau\rightarrow 0$. It is obtained by matching the resummed cross-section (first term in the r.h.s. of Eq. (\ref{eq:Rres}))
to the fixed-order one. The matching of the two contributions has to be carried out taking care of avoiding double counting of logarithms appearing in both of them.
To this purpose we follow the R and log-R schemes discussed in refs. \cite{CTTW,Jones:2003yv,Monni:2011gb}. The matched cross section and the resulting distribution have to vanish at the kinematical endpoint
$\tau_{\rm max}$. 
The resummed contribution does not trivially fulfill this property which has to be imposed in Eq.~(\ref{eq:csigma2}) by replacing the resummed logarithms
according to
\begin{align}
L\rightarrow \frac{1}{p}\ln\left(\frac{1}{(x_{L}\tau)^p}-\frac{1}{(x_{L}\tau_{\rm max})^p}+1\right)\,,\label{eq:log_shift}
\end{align}
where $p$ is the modified-logarithm parameter, whose default value is set to $p=1$. Furthermore in the R scheme the single logarithm coefficients $G_{n1}$ have to be modified as
follows:
\begin{align}
\label{eq:Gn1-shift}
 G_{n1}\rightarrow G_{n1}\left[1-\left(\frac{\tau}{x_{L}\tau_{\rm max}}\right)^p\right]\,.
\end{align}
Eqs.~(\ref{eq:log_shift},\ref{eq:Gn1-shift}) also introduce the resummation scale $x_L$ which accounts for the ambiguity in defining the resummed logarithms. The resummation functions $g_i(\alpha_{s}L)$ as well as the
coefficients $G_{nm}$ and $C_{k}$ also acquire a dependence on $x_L$ and need to be modified accordingly~\cite{Jones:2003yv,Monni:2011gb}.

In view of matching the NNLL resummed distribution to the NNLO fixed
order prediction (\ref{eq:Rfixedorder}) using the $R$-matching scheme, we need to include the logarithmically subleading terms $C_{2}$, $C_{3}$
and $G_{31}$ in the expansions (\ref{eq:csigma1}) and (\ref{eq:csigma2}).
Eq.~(\ref{eq:Rres}) then produces a reliable prediction of the thrust distribution over a broader kinematical range. 

The thrust cross section at NLL accuracy was computed in~\cite{CTTW}. The resummation beyond this accuracy was first achieved in \cite{Becher:2008cf} using an effective-theory approach and revisited
in \cite{Monni:2011gb},
where the full analytic expressions for the $\mathcal{O}(\bar{\alpha}_{s}^2)$ constant term $C_{2}$ and the coefficient $G_{31}$ were also obtained.
The $\mathcal{O}(\bar{\alpha}_{s}^3)$ constant term $C_{3}$ is currently unknown, and a numerical estimate is given in \cite{Monni:2011gb} together with the analytic expressions of
the functions $g_{i}(\alpha_{s}L)$. We report a summary of the resummation formulae in~\ref{sec:resummformul}.

To compute the functions $g_{i}(\alpha_{s}L)$, we recast the cross-section (\ref{eq:Rfixed}) in the dijet limit as
\begin{align}\label{eq:Rlaplace}
R_{T}(\tau)\,=\,\frac{1}{2\pi i}\int_{C} \frac{dN}{N}e^{\tau N} \tilde{\sigma}_{N}(Q^{2},\alpha_{s})+\mathcal{O}(\tau),
\end{align}
where ($N_{0}=e^{-\gamma_{E}}$)
\begin{align}
\tilde{\sigma}_{N}(Q^{2},\alpha_{s})\,=\,&H(1,\alpha_{s}(Q))\notag\\
\label{resummedcross}
\times\tilde{J}^{2}\left(1,\alpha_{s}(\sqrt{\frac{N_{0}}{N}}Q)\right)&\tilde{S}\left(1,\alpha_{s}(\frac{N_{0}Q}{N})\right)e^{-R\left(\frac{N_0}{N}\right)}\,.
\end{align}

The radiator $R({N_0}/{N})$ encodes the probability to emit a dressed gluon ({\it i.e.} with radiative corrections)  with momentum
\begin{align}
\label{eq:parametrisation}
k = u p + v \bar{p} + k_\perp,
\end{align}
where $p$ and $\bar{p}$ are the momenta of the back-to-back quark and antiquark, respectively.
Its expression reads
\begin{align}
\label{eq:radiator}
R\left(\frac{N_0}{N}\right) = 2\int_{\frac{N_{0}}{N}}^{1}\frac{du}{u}\int_{u^{2}Q^{2}}^{uQ^{2}}&\frac{dk_\perp^{2}}{k_\perp^{2}}\mathcal{A}(\alpha_{s}(k_\perp^{2}))\notag\\
&+2\int_{\frac{N_{0}}{N}}^{1}\frac{du}{u}\mathcal{B}(\alpha_{s}(uQ^{2}))\,.
\end{align}
The function $\mathcal{A}(\alpha_{s})$ has a soft origin, while $\mathcal{B}(\alpha_{s})$ accounts for the hard collinear contributions. They both admit a perturbative expansion in $\alpha_s$ 
\begin{align}
\mathcal{A}(\alpha_{s}) = \sum_i\frac{\alpha_s^i}{\pi^i}A^{(i)}, \qquad \mathcal{B}(\alpha_{s}) = \sum_i\frac{\alpha_s^i}{\pi^i}B^{(i)},
\end{align} 
and their expressions are reported in~\ref{sec:resummformul}. 
In the previous formulae we systematically neglected terms of order $\mathcal{O}(\tau)$ which are absorbed in the remainder function $D(\tau,\alpha_{s})$.
The integration countour in Eq. (\ref{eq:Rlaplace})
runs parallel to the imaginary axis to the right of all singularities of the integrand. For a description of the terms in Eq.~(\ref{resummedcross})  (each of which admits an expansion in $\alpha_{s}$)
we refer to \cite{Becher:2008cf,Monni:2011gb}. \\
From Eq.~(\ref{eq:radiator}) we see that the integral over $u$ in the exponent is regularised by the lower bound $\frac{N_{0}}{N}$. Such a bound acts as
an infrared regulator which prevents the strong coupling constant from being evaluated at non-perturbative scales ($\leq \Lambda_{QCD}$). This can be seen from Eq. (\ref{eq:Rlaplace})
where the contour should be set away from all the singularities (in particular from the Landau pole). Nevertheless, for resummation purposes we can set the contour to the left
of the Landau singularity since it would contribute with a non-logarithmic effect suppressed by some negative power of the centre-of-mass energy scale.
Such contributions, together with hadronisation corrections, are numerically sizeable at low energies and they are relevant to fit the strong coupling. \\
Several phenomenological hadronisation models
can
be found in the literature: analytical~\cite{Dokshitzer:1997ew, shapefunction, dressedgluon, Hoang:2007vb},
empirical~\cite{pythia,herwig,sherpa} and statistical~\cite{theFlorenceGroup}. Each of them is based on different theoretical assumptions.
In our
analysis we restrict ourselves to the analytical dispersive model developed in~\cite{Dokshitzer:1995qm,Dokshitzer:1997ew}. 

\section{Finite bottom-quark mass corrections}
\label{sec:bottom-mass corrections}
An additional effect to be considered concerns the assumption of vanishing quark masses we made in computing the cross section.
Such an assumption is not fully justified for low energy data samples ({\it e.g.} LEP1, PETRA energies)
for which finite bottom quark mass effects are relevant at the percent level~\cite{quarkmass, nasonoleari}.
We include bottom mass corrections directly at the level of the matched distribution by subtracting the fraction of massless $b$-quark events ($r_{b}(Q)$) and adding back the corresponding massive
contributions. This results in
\begin{align}
\frac{1}{\sigma}\,\frac{\d\sigma}{\d \tau}(\tau,Q) = \frac{1}{\sigma}&\left((1-r_{b}(Q))\frac{\d\sigma}{\d \tau}(\tau,Q)|_{\textrm{massless}}+\right.\notag\\
&\left.+r_{b}(Q)\frac{\d\sigma}{\d \tau}(\tau,Q)|_{\textrm{massive}}\right).
\end{align}
Here $\frac{\d\sigma}{\d \tau}(\tau,Q)|_{\textrm{massless}}$ is the NNLL+NNLO matched distribution and $\frac{\d\sigma}{\d \tau}(\tau,Q)|_{\textrm{massive}}$ is the NLO massive distribution obtained
with the parton level Monte Carlo {\tt Zbb4}~\cite{nasonoleari}. Since the NNLO correction to the massive distribution is currently unknown, we replace it with the corresponding massless result
\begin{equation}
\frac{\d C}{\d \tau}(\tau)|_{\rm massive} = \frac{\d C}{\d \tau}(\tau)|_{\rm massless}. 
\end{equation}
The $b$-quark mass corrections are generated for the considered energies. Futhermore, we perform a bin-by-bin interpolation of the energy dependence in order to compute them for any value of
$Q$ included in the considered energy range.

\section{Non-perturbative corrections}
\label{sec:Non-perturbative corrections}

In the dispersive model~\cite{Dokshitzer:1995qm,Dokshitzer:1997ew} non-perturbative corrections are accounted for by means of an effective coupling $\alpha_{{\rm eff}}(k^2)$ which is supposed to be
finite in the infrared region down to $k^2\rightarrow 0$. 
To introduce the non-perturbative correction, we use the parametrisation (\ref{eq:parametrisation}) for the momentum of a soft gluon, or an object with gluon quantum numbers ({\it i.e.} the offspring
of a gluon decay).
The two {\it Sudakov} parameters $u$ and $v$ are related by $v = (k_\perp^2+m^2)/u$, where $m^2$ is the invariant mass of the parent gluon.
The strong coupling is defined in a physical scheme as the anomalous dimension $\Gamma$ controlling the evolution of the {\it inclusive} soft gluon emission probability
\begin{align}
dw = \frac{2 C_F}{\pi}\frac{du}{u}\frac{dk_{\perp}^2}{k_{\perp}^2}\Gamma(\alpha_s).
\end{align}
The physical expression of the coupling then becomes
\begin{align}
\label{eq:dispersive-relation-1}
\tilde{\alpha}_s(k_\perp^2) \equiv \Gamma(\alpha_s) = \tilde{\alpha}_s(0)+\int_0^\infty dm^2\frac{k_\perp^2}{k_\perp^2+m^2}\frac{d\alpha_{\rm eff}(m^2)}{dm^2},
\end{align}
where the {\it effective} coupling $\alpha_{\rm eff}$ is defined as the logarithmic derivative of the spectral density for the dispersive relation (\ref{eq:dispersive-relation-1}).
Integrating (\ref{eq:dispersive-relation-1}) by parts (notice that $\tilde{\alpha}_s(0) = \alpha_{\rm eff}(0)$) we obtain a dispersive relation for the physical coupling $\tilde{\alpha}_s$
\begin{align}
\label{eq:dispersive-relation}
\tilde{\alpha}_s(k_\perp^2) = \int_0^\infty dm^2\frac{k_\perp^2}{(m^2+k_\perp^2)^2}\alpha_{\rm eff}(m^2).
\end{align}
This relation ensures that the strong coupling constant $\tilde{\alpha}_s(k_\perp^2)$ remains finite at very low momenta, where the perturbative picture fails.
It can be recast symbolically as the sum of a perturbative and a non-perturbative components
\begin{align}
\label{eq:alpha-split}
\tilde{\alpha}_s(k_\perp^2) = \alpha_s^{\rm PT}(k_\perp^2)+\alpha_s^{\rm NP}(k_\perp^2).
\end{align}
None of the two terms in the r.h.s. of~(\ref{eq:alpha-split}) is separately well defined in the infrared region, but their sum is finite down to $k_\perp^2=0$. 

The perturbative component in Eq.~(\ref{eq:alpha-split}) of the physical coupling is defined as the perturbative contribution to the physical scheme~\cite{Catani:1990rr} used above to define the
strong coupling $\tilde{\alpha}_s$.
The inclusive-probability picture used in this scheme requires that we integrate {\it inclusively} over the invariant mass $m^2$ of the parent gluon. Such an approximation is accurate up to NLL and to
this accuracy the relation between the physical and the $\overline{\rm MS}$ coupling can be extracted from the soft contribution to the Sudakov radiator~(\ref{eq:radiator}) and
reads~\cite{Catani:1990rr}
\begin{align}
\label{eq:CMW-alphas2}
\alpha_s^{\rm PT} = \alpha_s\left(1 + \frac{\alpha_s}{\pi} \frac{A^{(2)}}{A^{(1)}}+\dots\right).
\end{align} 
At NNLL, the non-inclusiveness of the thrust plays an important role, and it must be taken into account. We need to account for it both at the perturbative as well as at the non-perturbative level. 
At the perturbative level we can extend Eq.~(\ref{eq:CMW-alphas2}) by including the NNLL soft contribution in Laplace space (where the thrust factorises) $A^{(3)}$, which correctly contains non-inclusive effects
\begin{align}
\label{eq:CMW-alphas3}
\alpha_s^{\rm PT} = \alpha_s\left(1 + \frac{\alpha_s}{\pi} \frac{A^{(2)}}{A^{(1)}}+\frac{\alpha_s^2}{\pi^2}\frac{A^{(3)}}{A^{(1)}}+\dots\right).
\end{align}
It should be noted that the NNLL perturbative coefficient $A^{(3)}$ might be different for different observables. This would lead to a breakdown of the universality of the physical scheme for hadronisation corrections at this order.

\noindent According to Eq.~(\ref{eq:alpha-split}), the effective coupling $\alpha_{\rm eff}$ can be split into a perturbative term $\alpha_{\rm eff}^{\rm PT}$ and a non-perturbative contribution
$\alpha_{\rm eff}^{\rm NP}$ which is supposed to be a rapidly falling function in the ultra-violet region, where the running of the physical coupling~(\ref{eq:dispersive-relation}) matches the
solution of the renormalisation group equation for the perturbative component $\alpha_s^{\rm PT}$. The former is responsible for the perturbative strong coupling $\alpha_s^{\rm PT}$ in
Eq.~(\ref{eq:alpha-split}), while the latter gives rise to the non-perturbative component $\alpha_s^{\rm NP}$. 
Imposing that for large momenta $\alpha_s^{\rm NP}$ does not produce any power correction leads to the relation~\cite{Dokshitzer:1995qm,Shifman:1978bx}\\
\begin{align}
\label{eq:itep-ope}
\int_0^\infty \frac{dm^2}{m^2} m^{2 n}\alpha_{\rm eff}^{\rm NP}(m^2) = 0,\qquad n=1,2,...
\end{align}

By means of Eq.~(\ref{eq:dispersive-relation}) we can find the following useful relations between the perturbative and non-perturbative components of $\alpha_{\rm eff}$ and the respective couplings
\begin{align}
\label{eq:relation-pt}
\int_0^\infty\frac{dk_\perp^2}{k_\perp^2}\alpha_s^{\rm PT}(k_\perp^2) &= \int_0^\infty\frac{dm^2}{m^2}\alpha_{\rm eff}^{\rm PT}(m^2),\\
\label{eq:relation-nonpt}
\int_0^{\mu_I}dk_\perp\alpha_s^{\rm NP}(k_\perp^2) &= \frac{\pi}{4}\int_0^\infty\frac{dm^2}{m^2}m\,\alpha_{\rm eff}^{\rm NP}(m^2).
\end{align}
To obtain Eq.~(\ref{eq:relation-nonpt}) we integrated over $k_\perp^2$ and expanded the result in a power series of $m^2$. This is motivated by the fact that $\alpha_{\rm eff}^{\rm NP}(m^2)$ is
concentrated at small scales. Using Eq.~(\ref{eq:itep-ope}) it is straightforward to see that only the first term of the series survives yielding Eq.~(\ref{eq:relation-nonpt}).\\
In order to match the non-perturbative and the perturbative couplings we introduce an infrared matching scale $\mu_I\geq\Lambda_{\rm QCD}$ such that $\alpha_s^{\rm NP}(k^2)$ is negligible for $k^2\geq \mu_I^2$,
and the coupling is well approximated by the perturbative part $\alpha_s^{\rm PT}$. 
Since we are only interested in the leading power correction ($\sim 1/Q$), we limit our analysis to the region where $\tau\gg\mu_{I}/{Q}$ and we assume that the effective coupling is small
enough to neglect terms of order $\mathcal{O}(\alpha_{\rm{eff}}^{2})$. It is straightforward to see that all the terms but $R(N_0/N)$ in Eq.~(\ref{resummedcross}) are evaluated
at perturbative scales ({\it i.e.} the scales at which the strong coupling is evaluated are always larger than the matching scale $\mu_{I}$).
This implies that the only contribution to the leading power correction arises from the radiator
$R(N_0/N)$. In Eq.~(\ref{eq:radiator}) the latter is expressed in terms of the perturbative coupling in the $\overline{\rm MS}$ scheme and we need to redefine it using the full physical
coupling~(\ref{eq:dispersive-relation}).
 We first rewrite Eq.~(\ref{eq:radiator}) using the identity~\cite{Catani:2003zt}
\begin{align}\label{eq:catanirelation}
e^{-N u}-1 &=-\Theta\left(u-\frac{N_{0}}{N}\right)\nonumber\\
&+\Gamma_{2}(\partial_{\ln N})\partial^{2}_{\ln N}\Theta\left(u-\frac{N_{0}}{N}\right)+\mathcal{O}\left(\frac{N_{0}}{N}\right),
\end{align}

\noindent where
\begin{align}
\Gamma_{2}(\epsilon) &= \frac{1}{\epsilon^{2}}(1-e^{-\gamma_{E}\epsilon}\Gamma(1-\epsilon))\nonumber\\
&=-\frac{\zeta_{2}}{2}-\frac{\zeta_{3}}{3}\epsilon+\mathcal{O}(\epsilon^{2}),
\end{align}
and the derivative is meant to act on the whole integral whose boundaries are set by the $\Theta(u-N_{0}/N)$ function.
Eq.~(\ref{eq:radiator}) then becomes
\begin{align}
\label{eq:modified-radiator}
&R\left(\frac{N_0}{N}\right) = -2\int_{0}^{1}\frac{du}{u}(e^{-N\,u}-1)\int_{u^{2}Q^{2}}^{uQ^{2}}\frac{dk_\perp^2}{k_\perp^2}\mathcal{A}(\alpha_{s}(k_\perp^{2}))\nonumber\\
&\qquad\, -2\int_{0}^{1}\frac{du}{u}(e^{-N\,u}-1)\mathcal{B}(\alpha_{s}(uQ^{2}))\notag\\
&-\zeta_2\bigg(2\mathcal{A}\left(\alpha_{s}\left(\frac{N_{0}^2}{N^2}Q^2\right)\right)-\mathcal{A}\left(\alpha_{s}\left(\frac{N_{0}}{N}Q^2\right)\right)\notag\\
&\qquad\, +\partial_{\ln N}\mathcal{B}(\alpha_{s}(\frac{N_0}{N}Q^{2}))\bigg) +\mathcal{O}\left(\alpha_s^n\ln^{n-2} N\right) \,,
\end{align}
where we used the Leibniz's rule to perform the derivatives of the integrals. 
It is straightforward to check that the term proportional
to $\zeta_2$ contributes to NNLL accuracy. However, we see that it is always evaluated at perturbative scales, so it does not contribute to the leading power correction. 
We now observe that the integral involving the collinear $\mathcal{B}(\alpha_{s}(uQ^{2}))$  function in Eq.~(\ref{eq:modified-radiator}) gives rise to a subleading power-suppressed term which scales as
$1/Q^2$, so we can neglect it in the non-perturbative analysis.
The only contribution to the leading power correction stems from the double integral in Eq.~(\ref{eq:modified-radiator}), involving the soft contribution $\mathcal{A}(\alpha_s(k_\perp^2))$.
Making use of Eq.~(\ref{eq:CMW-alphas3}) we can recast this contribution using the physical coupling $\alpha_s^{\rm PT}$ (to NNLL accuracy) as

\begin{align}
\label{eq:radiator-nonpt}
R\left(\frac{N_0}{N},\alpha_s\right) &= -\frac{2C_F}{\pi}\int_0^1\frac{du}{u}(e^{-N\,u}-1)\int\frac{dk_\perp^{2}}{k_\perp^{2}}\alpha_s^{\rm PT}(k_\perp^2)\notag\\
&\times\Theta(k_\perp^2-u^2Q^2)\Theta(uQ^2-k_\perp^2)+\dots,
\end{align}
where the ellipsis stand for the remaining perturbative terms in the radiator. Eq.~(\ref{eq:radiator-nonpt}) is nothing but the single soft emission contribution to the cross section, with the coupling being
defined in the physical scheme.\\
To correctly account for non-perturbative corrections, we replace $\alpha_s^{\rm PT}$ in Eq.~(\ref{eq:radiator-nonpt}) with the full coupling~(\ref{eq:dispersive-relation}). Moreover, since the
physical definition of the coupling deals with {\it massive} gluons, the massless phase space in Eq.~(\ref{eq:radiator-nonpt}) is modified to take into account the gluon mass. This amounts to
performing the replacement $k_\perp^2\rightarrow k_\perp^2+m^2$ in the two $\Theta$-functions in Eq.~(\ref{eq:radiator-nonpt}), obtaining
\begin{align}
\label{eq:radiator-nonpt-massive}
&R\left(\frac{N_0}{N},\alpha_s\right) = -\frac{2C_F}{\pi}\int_0^1\frac{du}{u}(e^{-N\,u}-1)\notag\\
&\times\int\frac{dk_\perp^{2}}{k_\perp^{2}}\left(\tilde{\alpha}_s(0)+\int_0^\infty dm^2\frac{k_\perp^2}{k_\perp^2+m^2}\frac{d\alpha_{\rm eff}(m^2)}{dm^2}\right)\notag\\
&\times\Theta(k_\perp^2+m^2-u^2Q^2)\Theta(uQ^2-k_\perp^2-m^2)+\dots
\end{align}
We integrate by parts over $m^2$ in  Eq.~(\ref{eq:radiator-nonpt-massive}) getting rid of the boundary term $\tilde{\alpha}_s(0)$, and we integrate over $k_\perp^2$:
\begin{align}
&R\left(\frac{N_0}{N},\alpha_s\right) =
 -\frac{2C_F}{\pi}\int_0^1\frac{du}{u}(e^{-N\,u}-1)\notag\\
&\int_0^\infty \frac{dm^2}{m^2}\alpha_{\rm eff}(m^2)\Theta(m^2-u^2Q^2)\Theta(uQ^2-m^2)+\dots
\end{align} 
We now split the effective coupling into its perturbative ($\alpha_{\rm eff}^{\rm PT}$) and non-perturbative ($\alpha_{\rm eff}^{\rm NP}$) components.

For the perturbative term we use Eq.~(\ref{eq:relation-pt}) to reproduce the perturbative soft piece of the radiator~(\ref{eq:radiator-nonpt}). Replacing the perturbative coupling $\alpha_s^{\rm PT}$
with its expression (\ref{eq:CMW-alphas3}) we reproduce Eq.~(\ref{eq:modified-radiator}). We then can write the radiator as 

\begin{align}
\label{eq:radform}
R\left(\frac{N_0}{N},\alpha_s\right) = R^{\rm PT}\left(\frac{N_0}{N},\alpha_s\right)+R^{\rm NP}\left(\frac{N_0}{N},\alpha_s\right).
\end{align}
$R^{\rm PT}(\frac{N_0}{N},\alpha_s)$ is the full perturbative radiator~(\ref{eq:modified-radiator}) that we recast as
\begin{align}
R^{\rm PT}\left(\frac{N_0}{N},\alpha_s\right) &= -\ln N h_1(\as\ln N)
-h_2(\as\ln N)\notag\\
&-\frac{\alpha_{s}}{\pi}\beta_{0}h_3(\as\ln N)+\mathcal{O}(\alpha_s^n\ln^{n-2} N),
\end{align}
where the functions $h_i(\alpha_s \ln N)$ are reported in~\ref{sec:resummformul}.
The second term in the r.h.s. of Eq.~(\ref{eq:radform}) $R^{\rm NP}(\frac{N_0}{N},\alpha_s)$ is the non-perturbative component, given by
\begin{align}
&R^{\rm NP}\left(\frac{N_0}{N},\alpha_s\right) = -\frac{2C_F}{\pi}\int_0^1\frac{du}{u}(e^{-N\,u}-1)\notag\\
&\times\int_0^\infty \frac{dm^2}{m^2}\alpha_{\rm eff}^{\rm NP}(m^2)\Theta(m^2-u^2Q^2)\Theta(uQ^2-m^2).
\end{align}
The non-perturbative term requires some more attention. We first recall that we are working in the approximation $\mu_I/Q\ll \tau$, which allows one to expand the exponential function as
$e^{-N\,u}\simeq 1-Nu+...$, neglecting subleading terms since they give rise to $\mathcal{O}(1/Q^2)$ corrections. It is then straightforward to perform the integral over $u$, obtaining 
\begin{align}
\label{eq:nonpt-radiator}
R^{\rm NP}\left(\frac{N_0}{N},\alpha_s\right) = N\frac{2C_F}{\pi}\int_0^\infty \frac{dm^2}{m^2}\alpha_{\rm eff}^{\rm NP}(m^2)\left(\frac{m}{Q}-\frac{m^2}{Q^2}\right).
\end{align}
The term proportional to $m^2$ in the round brackets leads to a vanishing contribution because of Eq.~(\ref{eq:itep-ope}). Making use of Eq.~(\ref{eq:relation-nonpt}) we end up with
\begin{align}
\label{eq:final-nonpt}
R^{\rm NP}\left(\frac{N_0}{N},\alpha_s\right) = \frac{N}{Q}\frac{8C_F}{\pi^2}\int_0^{\mu_I} dk_\perp\alpha_{s}^{\rm NP}(k_\perp^2).
\end{align}
To evaluate Eq.~(\ref{eq:final-nonpt}) we replace $\alpha_s^{\rm NP}=\tilde{\alpha}_s-\alpha_s^{\rm PT}$. Introducing the mean value of the physical coupling below $\mu_I$ 
\begin{align}
\alpha_0(\mu_I) = \frac{1}{\mu_I}\int_0^{\mu_I}dk_\perp\tilde{\alpha}_s(k_\perp^2),
\end{align}
and expanding perturbatively $\alpha_s^{\rm PT}$ to perform the integral, we obtain
\begin{align}
R^{\rm NP}\left(\frac{N_0}{N},\alpha_s\right) = N\Delta\tau.
\end{align}
Using the expression (\ref{eq:CMW-alphas3}) for $\alpha_s^{\rm PT}$, the quantity $\Delta\tau$ amounts to
\begin{align}
\label{eq:shift}
\Delta\tau =& \frac{\mu_{I}}{Q}\frac{8C_F}{\pi^2}\bigg[\alpha_{0}(\mu_{I}^{2})-\alpha_{s}(\mu_{R}^2)\nonumber\\
&-\alpha_{s}^{2}(\mu_{R}^2)\frac{\beta_{0}}{\pi}\left(2\ln\frac{\mu_{R}}{\mu_{I}}+\frac{A^{(2)}}{A^{(1)}\beta_{0}}+2\right)\nonumber\\
&-\alpha_{s}^{3}(\mu_{R}^2)\frac{\beta_{0}^2}{\pi^2}\bigg(4\ln^{2}\frac{\mu_{R}}{\mu_{I}}+4\bigg(\ln\frac{\mu_{R}}{\mu_{I}}+1\bigg)\nonumber\\
&\times\bigg(2+\frac{\beta_{1}}{2\beta_{0}^2}+\frac{A^{(2)}}{A^{(1)}\beta_{0}}\bigg)+\frac{A^{(3)}}{A^{(1)}\beta_{0}^2}\bigg)\bigg].
\end{align}

Note that the NNLO contribution $A^{(3)}$ in Eq.~(\ref{eq:shift}) is found to be different from what obtained in \cite{jaquier} and \cite{davisonwebber}, where two different assumptions
for this new term were made. 
The final expression for $A^{(3)}$~\cite{Monni:2011gb} is reported in~\ref{sec:resummformul} and
it consists of the sum of two contributions: the observable-independent three-loop cusp anomalous dimension computed in~\cite{Vogt:2004mw} and an observable-dependent term proportional to the two-loop soft anomalous dimension obtained from the
${\cal O}(\alpha_s^2)$ soft contribution to the thrust cross section~\cite{Monni:2011gb,Becher:2008cf,Kelley:2011ng,Hornig:2011iu}. The latter turns out to give the leading numerical contribution to
$A^{(3)}$.\\
Using Eq.~(\ref{eq:shift}) in Eq.~(\ref{eq:Rlaplace}) results in a shift of the cross section by an amount $\Delta\tau$. This shift encodes the leading non-perturbative correction to the thrust cross
section.\\
The result in Eq.~(\ref{eq:shift}) is not yet complete. So far we have considered the dispersive model in its inclusive form. The non-perturbative effect of the thrust's non-inclusiveness can be accounted for using
perturbation theory by computing the correction to the form reported in Eq.~(\ref{eq:final-nonpt}). Since the physical coupling is defined as the soft emission probability, one can compute the
corrections due to incomplete cancellations between real and virtual contributions as well as to scenarios in which the progeny of the massive gluon goes into 
opposite hemispheres. These corrections were computed up to $\mathcal{O}(\alpha_s^2)$ in~\cite{Dokshitzer:1997iz} where it was shown that they amount to a multiplicative ({\it Milan}) factor ${\cal
M}$
\begin{align}\label{eq:milan}
\mathcal{M}=1+\frac{1}{4\beta_{0}}(1.575\,C_{A}-0.104\,n_{F})= 1.490, \,\,\,n_{F}=3,
\end{align}
where we set the number of active flavours to 3 since it is only sensitive to low energy soft radiation. In particular, the $n_{F}$ factor in (\ref{eq:milan})
is due to a soft gluon splitting into a $q\bar{q}$ pair of light quarks.
It is difficult to estimate the uncertainty on ${\cal M}$. It can not be excluded that higher order $\mathcal{O}(\alpha_s^3)$ corrections to the Milan factor could be as large as
$20\%$~\cite{Dokshitzer:1998nz}. We use this value as uncertainty on ${\cal M}$ in our analysis.
Since the matched distribution is given in terms of a binned histogram, the shift~(\ref{eq:shift}) cannot be straightforwardly performed since in general $\Delta\tau$ is not a multiple of the bin
width. We then interpolate between different bins with a cubic spline with the ``not-a-knot'' condition in order to evaluate the distribution at intermediate values of $\tau$. The
shift~(\ref{eq:shift}) is then performed directly on the resulting interpolating spline.

\subsection{\bf Hadron masses and decay effects}

In deriving the dispersive contribution to the leading power  
correction ($\sim1/Q$), all offspring particles produced in the {\it massive} gluon decay were assumed to be massless.
Taking their finite masses into account would change the expression of  
the thrust in the soft region, leading to an additional kind of power  
corrections~\cite{Salam:2001bd} whose leading scaling is $\ln^A Q/Q$,  
where $A$ is a constant. Such corrections are present in experimental data, where massive hadrons contribute to the thrust. Moreover, a similar power-suppressed term arises from the final state
momenta reshuffle due to unstable hadron decays into lighter particles.
Unlike the dispersive contribution, these power correction cannot be expressed as the product
of a universal non-perturbative quantity ({\it e.g.} $\alpha_0$) and  
an observable dependent factor, thus they break universality.
A direct consequence is that they have a completely different  
impact on different event shape observables.
An initial detailed assessment of hadron mass effects on event  
shapes has been presented in~\cite{Salam:2001bd}, where the constant $A$ was evaluated under the hypothesis of local parton-hadron duality, and assuming  
massless partons in the perturbative calculation, finding $A\,=\,4C_A/ 
\beta_0$. Bottom quark mass effects in the perturbative contribution could potentially modify this value. A more recent systematic analysis of hadron mass effects on power corrections has been performed in Ref.~\cite{Mateu:2012nk}.
In Refs.~\cite{Salam:2001bd,Mateu:2012nk} it is shown that the universality of mass-dependent power corrections to two-jet event-shape cross sections can be {\it rescued} by properly redefining the measurement scheme. This allows one to study systematically the impact of non-perturbative mass corrections
for different event-shape observables. 
In particular, they were found to lead to sizeable corrections for jet-masses and related quantities, while their impact on the thrust is
very much limited~\cite{Salam:2001bd,Mateu:2012nk}. Consequently, we do not include hadron-mass effects in our study here.



\section{Determination of $\alpha_s$ and $\alpha_0$}
\label{sec:Determination of alpha}
In order to measure the strong coupling constant $\alpha_s$ 
and the non-perturbative parameter $\alpha_0$ we fit the QCD predictions to experimental data reported in Table~\ref{tab:data}, in the centre-of-mass energy range ${\rm 14\,GeV} \leq Q \leq {\rm
206\,GeV}$. For consistency we consider data sets which have been corrected for QED radiation effects. The other data sets would require the inclusion of electroweak corrections~\cite{Denner:2009gx} into the theoretical description. Therefore we limit our analysis to data measured by the ALEPH~\cite{aleph},
the L3~\cite{l3} and the TASSO~\cite{tasso} collaborations.
\begin{table}
\begin{tabular}{|l|r|c|r|c|}
  \hline
  Exp. & Q (GeV) & Fit range  & N.~Pts. & Ref. \\
  \hline
  TASSO      & 14.0    &  $0.200<\tau<0.32$  &  3 & \cite{tasso} \\
  TASSO      & 22.0    &  $0.120<\tau<0.32$  &  5 & \cite{tasso} \\
  TASSO      & 35.0    &  $0.080<\tau<0.32$  &  7 & \cite{tasso} \\
  TASSO      & 44.0    &  $0.080<\tau<0.32$  &  7 & \cite{tasso} \\
  ALEPH      & 91.2    &  $0.050<\tau<0.33$  & 28 & \cite{aleph} \\
  ALEPH      & 133.0   &  $0.040<\tau<0.30$  &  7 & \cite{aleph} \\
  ALEPH      & 161.0   &  $0.040<\tau<0.30$  &  7 & \cite{aleph} \\
  ALEPH      & 172.0   &  $0.040<\tau<0.30$  &  7 & \cite{aleph} \\
  ALEPH      & 183.0   &  $0.040<\tau<0.30$  &  7 & \cite{aleph} \\
  ALEPH      & 189.0   &  $0.040<\tau<0.30$  &  7 & \cite{aleph} \\
  ALEPH      & 200.0   &  $0.040<\tau<0.30$  &  7 & \cite{aleph} \\
  ALEPH      & 206.0   &  $0.040<\tau<0.30$  &  7 & \cite{aleph} \\
  L3         &  41.4   &  $0.065<\tau<0.33$  &  8 & \cite{l3}    \\
  L3         &  55.5   &  $0.065<\tau<0.33$  &  8 & \cite{l3}    \\
  L3         &  65.4   &  $0.065<\tau<0.33$  &  8 & \cite{l3}    \\
  L3         &  75.7   &  $0.045<\tau<0.33$  &  9 & \cite{l3}    \\
  L3         &  82.3   &  $0.045<\tau<0.33$  &  9 & \cite{l3}    \\
  L3         &  85.1   &  $0.045<\tau<0.33$  &  9 & \cite{l3}    \\
  L3         & 130.0   &  $0.050<\tau<0.30$  & 10 & \cite{l3}    \\
  L3         & 136.0   &  $0.050<\tau<0.30$  & 10 & \cite{l3}    \\
  L3         & 161.0   &  $0.050<\tau<0.30$  & 10 & \cite{l3}    \\
  L3         & 172.0   &  $0.050<\tau<0.30$  & 10 & \cite{l3}    \\
  L3         & 183.0   &  $0.050<\tau<0.30$  & 10 & \cite{l3}    \\
  L3         & 189.0   &  $0.050<\tau<0.30$  & 10 & \cite{l3}    \\
  L3         & 194.0   &  $0.050<\tau<0.30$  & 10 & \cite{l3}    \\
  L3         & 200.0   &  $0.050<\tau<0.30$  & 10 & \cite{l3}    \\
  L3         & 206.0   &  $0.050<\tau<0.30$  & 10 & \cite{l3}    \\
  \hline
\end{tabular}
    \caption{Data set considered for the simultaneous $\chi^{2}$ fit of $\alpha_s$ and $\alpha_0$.}
    \label{tab:data}
\end{table}

\begin{figure}
  \includegraphics[width=0.96\linewidth]{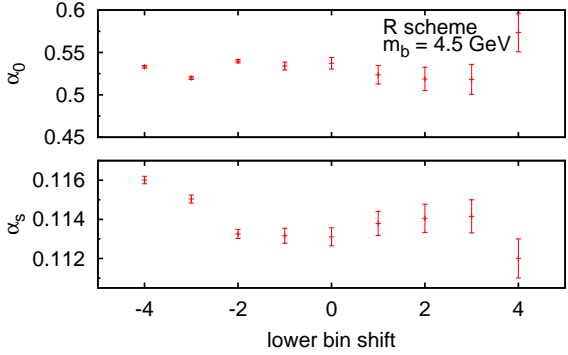}\\
  \caption{Check of the fit stability, when the lower bound of the fit range is shifted. The quoted error bars represent the statistical uncertainty on the single fit. The plots show that the chosen default range is in the middle of a stable plateau. Shifting the lower bound of
the fit to include more bins in the far infrared region leads to large deviations from the stable value of $\alpha_s$.}
  \label{fig:stability}
\end{figure}

 The theory distributions are computed in form of binned histograms. We rebin the theory results in order to match the data binning, which is
different for each experiment.
We set the upper limit of the fit range to $\tau \leq 1/3$ (the leading order kinematical upper limit) excluding the multi-particle region where the theoretical prediction fails because of the small
number of final-state partons considered. The lower limit is set two bins away from $\tau=\mu_I/Q$ excluding the peak region where our treatment of power corrections is not accurate.
Fig.~\ref{fig:stability} shows results for different fits of the strong coupling, obtained by shifting the lower bin around the default configuration (labeled with $0$). For values of the lower bin
shift below $-2$ the peak region is included in the fit range, which leads to higher values of $\alpha_{s}$. This region is excluded since the assumption $\tau\gg\mu_I/Q$ made in computing the leading
power correction breaks down and subleading contributions become numerically important.
The fit is performed by minimising the $\chi^2$ function defined as

\begin{align}
\chi^2 =& \sum_{i,j} \left(\left.\frac{1}{\sigma}\frac{d\sigma}{d\tau}(\tau_i)\right|^{\rm exp}-\left.\frac{1}{\sigma}\frac{d\sigma}{d\tau}(\tau_i)\right|^{\rm th}\right)V^{-1}_{ij}\notag\\
&\times\left(\left.\frac{1}{\sigma}\frac{d\sigma}{d\tau}(\tau_j)\right|^{\rm exp}-\left.\frac{1}{\sigma}\frac{d\sigma}{d\tau}(\tau_j)\right|^{\rm th}\right)\, ,
\end{align}
where $V_{ij}$ are the covariances of the distribution between the bins $\tau_i$ and $\tau_j$.
The general form of the covariance matrix is $V_{ij}= S_{ij}+E_{ij}$, where $S_{ij}=\sigma_{{\rm stat},\,i}^2\delta_{ij}$ is the diagonal matrix of the (uncorrelated) statistical errors, while
$E_{ij}$ contains the experimental systematic convariances.
The diagonal entries of $E_{ii}=\sigma_{{\rm syst},i}^2$ are given by the systematic uncertainty on the $i$-th bin, while we need to make a plausible assumption on the form of the off-diagonal
elements.
We consider the minimal-overlap model, which defines $E_{ij}$ as
\begin{equation}
E_{ij} = {\rm min}\left(\sigma_{{\rm syst},i}^2,\sigma_{{\rm syst},j}^2\right).
\end{equation} 
The $\chi^2$ minimization is carried out with the {\tt TMinuit} routine distributed with {\tt ROOT} and the whole analysis was implemented in a {\tt C++} code.

To estimate the fit uncertainties, different approaches have been proposed in the literature ~\cite{Jones:2003yv,davisonwebber}.
We decide to perform a scan of the theoretical parameters involved in the calculation as follows:
\begin{itemize}
\renewcommand{\labelitemi}{$\bullet$}
\item the renormalisation scale fraction $x_\mu$ (default value $x_\mu=1$) is randomly varied between 1/2 and 2,
\item the resummation scale fraction $x_L$ (default value $x_L=1$) is randomly varied between 2/3 and 3/2,
\item the Milan factor $\mathcal{M}$ is randomly varied within its theoretical uncertainty ($20\,\%$),
\item the default value of the modified-logarithm parameter $p=1$ is replaced by $p=2$,
\item the default $b$-quark mass ($m_b=4.5$~GeV) is replaced by $m_b=4.0$~GeV and $m_b=5.0$~GeV.
\end{itemize}
The random scans for $x_{\mu}$, $x_{L}$ and $\mathcal{M}$ are performed with high statistics for each combination of $m_b$ and $p$ ($~3\times 10^4$ fits in total).
The final error on $\alpha_s$ and $\alpha_0$ is obtained as the envelope of such variations. Moreover, we consider two different matching schemes (R and log-R) and take the difference between the two
central values as systematic uncertainty. The quoted central values correspond to the fit results with the default parameter setting.\\

\begin{figure}[t]
\centering
\includegraphics[width=0.96\linewidth]{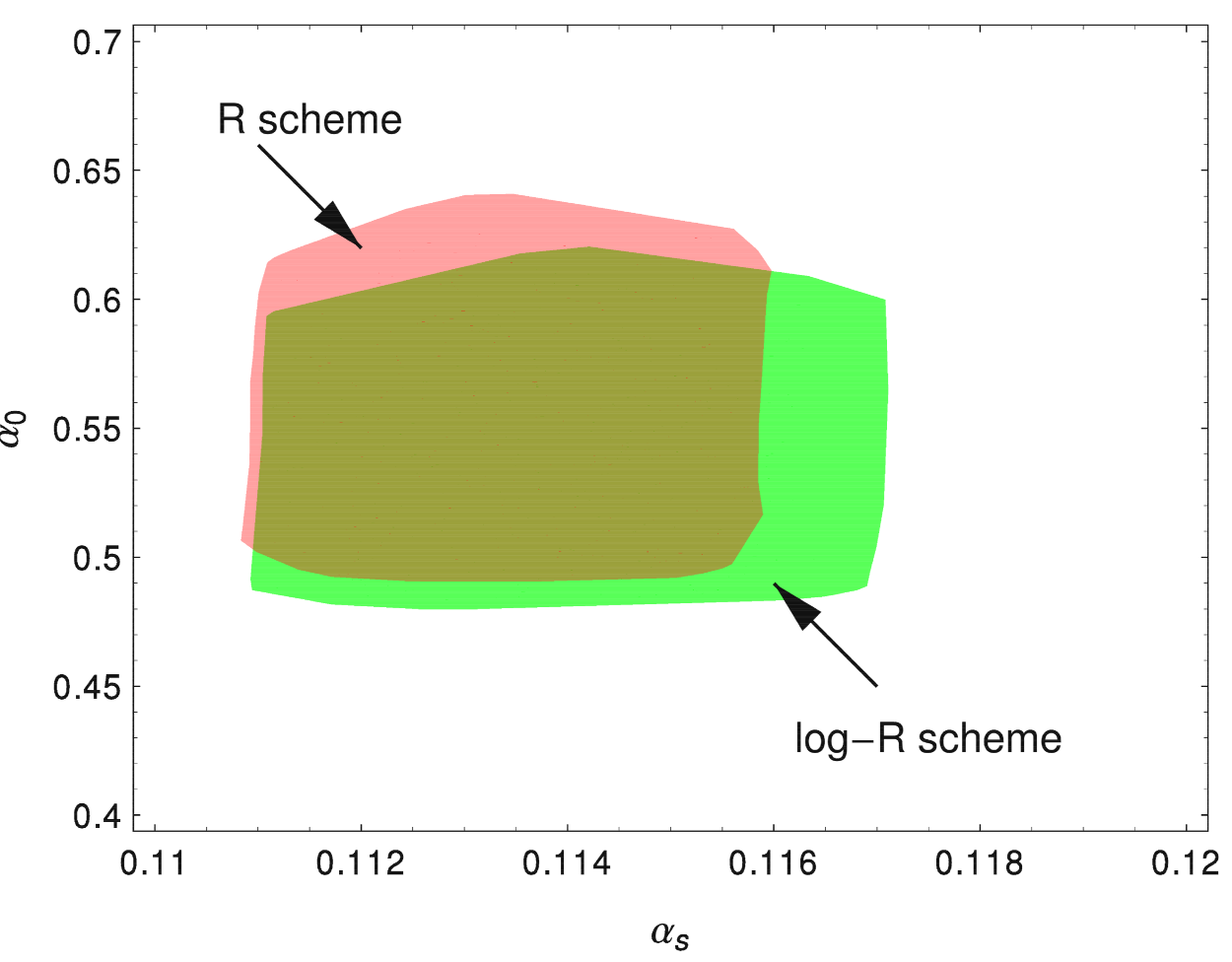}\hfill
\caption{Scatter plot for the simultaneous fit of $\alpha_s$ and $\alpha_0$ using two different matching schemes.}
\label{fig:NNLLresults}
\end{figure}

The scatter plots are shown in Fig.~\ref{fig:NNLLresults}. Using the  R scheme leads to a lower central value and a smaller uncertainty on $\alpha_s$ than in the log-R scheme.
We report the results separately for the R and log-R matching schemes:

\begin{center} 
\begin{tabular}{c|c|c|c}
{\rm scheme} & $\,\alpha_s(M_Z)$\, & $\,\alpha_0(2~{\rm GeV})$\, & $\chi^2$/d.o.f.\,
\\[0.2em]\hline
\phantom{x} & &
\\[-1em]
{\rm R} & $0.1131^{+0.0028}_{-0.0022}$ & $0.538^{+0.102}_{-0.047}$ \, & $1.54$\,
\\[0.4em]
{\rm log-R} & $0.1137^{+0.0034}_{-0.0027}$ & $0.524^{+0.096}_{-0.044}$\, & $1.60$\,
\end{tabular}
\end{center}

The difference between the central values is contained into the scan error bands, so it does not affect the final uncertainty. The final uncertainties on $\alpha_s$ and $\alpha_0$ can be considered to
a good  approximation as uncorrelated. This feature is visible from the scatter plot in Fig.~\ref{fig:NNLLresults}. To check the dependence of the $\alpha_s$ fit upon the infrared matching scale
$\mu_I$ we perform a fit replacing the default value $\mu_I=2$~GeV by $\mu_I=3$~GeV. The corresponding results for the mean effective coupling and for the perturbative strong coupling are:

\begin{center} 
\begin{tabular}{c|c|c|c}
{\rm scheme} & $\,\alpha_s(M_Z)$\, & $\,\alpha_0(3~{\rm GeV})$\, & $\chi^2$/d.o.f.\,
\\[0.2em]\hline
\phantom{x} & &
\\[-1em]
{\rm R} & $0.1130^{+0.0028}_{-0.0021}$ & $0.430^{+0.068}_{-0.032}$ \, & $1.56$\,
\\[0.4em]
{\rm log-R} & $0.1136^{+0.0034}_{-0.0027}$ & $0.422^{+0.060}_{-0.029}$\, & $1.62$\,
\end{tabular}
\end{center}

The $\alpha_s$ value is not affected by the infrared matching scale, whilst $\alpha_0$ decreases for larger values of $\mu_I$. We also observe that the relative uncertainty on the non-perturbative
parameter $\alpha_0$ is smaller for $\mu_I=3~{\rm GeV}$. This can be explained observing that the uncertainty on $\alpha_0$ is mostly controlled by the variation of the Milan factor ${\cal M}$. Given
the multiplicative relationship between the two quantities, a symmetric variation of ${\cal M}$ does not correspond to a symmetric uncertainty on $\alpha_0$. Moreover, it implies that for smaller
central values of $\alpha_0$, the size of the resulting uncertainty decreases. 
We have assumed that above $\mu_I = 2$~GeV the physical running coupling is well approximated by its perturbative component, this allows us to perform a consistency check on the running of $\alpha_0$. Since $\alpha_0$ is a mean value, we consider the integral of the physical $\tilde{\alpha}_s$ from $2$~GeV to $3$~GeV
\begin{align}
\int_{2\,{\rm GeV}}^{3\,{\rm GeV}}dk\,\tilde{\alpha}_s(k^2) = 3 \alpha_0(3\,{\rm GeV})-2 \alpha_0(2\,{\rm GeV}) = 0.214,
\end{align}
and we compare it to the perturbative result with four active flavours
\begin{align}
\int_{2\,{\rm GeV}}^{3\,{\rm GeV}}dk\,\alpha_s^{\rm PT}(k^2) = 0.218^{+0.013}_{-0.009},
\end{align}
where the values for $\alpha_0$ and $\alpha_s(M_Z)$ refer to the findings in the R matching scheme.
We observe that the two quantities are in very good agreement within the fit uncertainty. This observation indicates that the non-perturbative effects on the running of $\alpha_s$ can be considered negligible above $\mu_I=2$\,GeV, thereby justifying this choice for the non-perturbative matching scale.

To estimate the impact on $\alpha_s$ of the finite bottom-quark mass corrections we performed a further scan without including this correction (setting $m_b = 0$), using therefore only pure massless QCD results. With
this setting we obtain the following values

\begin{center} 
\begin{tabular}{c|c|c|c}
{\rm scheme} & $\,\alpha_s(M_Z)$\, & $\,\alpha_0(2~{\rm GeV})$\, & $\chi^2$/d.o.f.\,
\\[0.2em]\hline
\phantom{x} & &
\\[-1em]
{\rm R} & $0.1115^{+0.0022}_{-0.0017}$ & $0.539^{+0.122}_{-0.048}$ \, & $1.56$\,
\\[0.4em]
{\rm log-R} & $0.1123^{+0.0031}_{-0.0021}$ & $0.522^{+0.113}_{-0.047}$\, & $1.63$\,
\end{tabular}
\end{center}

The values for $\alpha_0$ remain almost unchanged while, with respect to the nominal scan,  the values of $\alpha_s$ in both matching schemes decrease. This can be explained by observing that the mass corrections tend to lower the thrust distribution in the fit range leading to a higher value for $\alpha_s$. The errors on $\alpha_s$ are slightly smaller since now we are not accounting for the uncertainty on $m_b$. 

\section{Comparison with other $\alpha_s$ determinations}
\label{sec:Comparison}
\begin{figure}
  \psfrag{dissnnlo}[cr][cr]{\scriptsize \cite{Dissertori:2007xa} (NNLO)}
  \psfrag{dissnnll}[cr][cr]{\scriptsize \cite{Dissertori:2009ik} (NLL+NNLO)}
  \psfrag{bethnnlo}[cr][cr]{\scriptsize \cite{jadeas} (NNLO)}
  \psfrag{bethnnll}[cr][cr]{\scriptsize \cite{jadeas} (NLL+NNLO)}
  \psfrag{opalnnlo}[cr][cr]{\scriptsize \cite{opalas} (NNLO)}
  \psfrag{opalnnll}[cr][cr]{\scriptsize \cite{opalas} (NLL+NNLO)}
  \psfrag{abbatexx}[cr][cr]{\scriptsize \cite{Abbate:2010xh} (N$^3$LL+NNLO)}
  \psfrag{bechschw}[cr][cr]{\scriptsize \cite{Becher:2008cf} (N$^3$LL+NNLO)}
  \psfrag{daviwebb}[cr][cr]{\scriptsize \cite{davisonwebber, Bethke:2011tr} (NLL+NNLO)}
  \psfrag{glmxxxxx}[cr][cc]{
	  \begin{minipage}{2cm}
          \centering
          \scriptsize This fit\\ (NNLL+NNLO)
          \end{minipage}}
  \psfrag{as}[cc][cc]{\scriptsize $\alpha_s$}
  \rule{-0.6cm}{0.0cm}\includegraphics[width=1.1\linewidth]{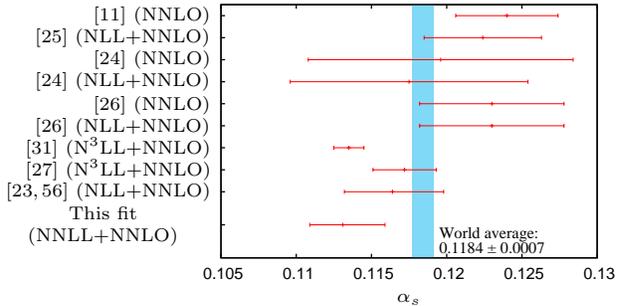}\\
  \caption{Comparison with other determinations of the strong coupling based on data for thrust distribution. The band shows the global world average~\cite{Bethke:2012zza}.}
\label{fig:othervalues}
\end{figure}

We compare our results to other recent determinations of the strong coupling based on thrust. The comparison is shown in Fig.~\ref{fig:othervalues}.
The central value is consistent with the findings of~\cite{Abbate:2010xh}. On the other hand our theory uncertainty is larger. The reduced uncertainty of~\cite{Abbate:2010xh} can be explained to some extent by including the subleading approximated N$^3$LL terms, which rely on a Pad\'e approximation for the four loop cusp anomalous dimension. Such terms carry a scale dependence which leads to a reduction of the final error. In our case, we do not include the full approximated N$^3$LL tower, but only consider the necessary term for the R matching scheme, {\it i.e.} $G_{31}\alpha_s^3
L$, for which an exact analytic expression is available~\cite{Monni:2011gb}. Moreover, a different non-perturbative model based on a shape function approach was used in~\cite{Abbate:2010xh}, and a
different technique to perform the renormalon subtraction was implemented. The latter turned out to have an important impact on the fit uncertainty.  \\

In the context of the dispersive model, a global fit was performed in~\cite{davisonwebber} to NLL+NNLO accuracy in the log-R matching scheme, using a larger data set. The quoted values for the perturbative coupling and
the non-perturbative parameter were updated in~\cite{Bethke:2011tr} and are $\alpha_s(M_Z) = 0.1164^{+0.0034}_{-0.0032}$ and $\alpha_0(2~{\rm GeV}) = 0.62\pm 0.02$, respectively. In the log-R matching scheme, we note that our resulting fit
uncertainty is comparable to what was obtained in~\cite{Bethke:2011tr}, despite the higher logarithmic accuracy of the resummation  in our analysis. 
This is because we consider the simultaneous variation of different parameters to assess the theory error, leading to a more conservative estimate. 
The uncertainty in the $\alpha_0$ value quoted above does not account for the $20\%$ variation of the Milan factor. Considering such a systematic uncertainty leads to a similar theoretical error on $\alpha_0$ ({\it cf.} erratum to~\cite{davisonwebber}).
For completeness, we report our fit result obtained with NLL$+$NNLO perturbative accuracy in the log-R scheme:

\begin{center} 
\begin{tabular}{c|c|c|c}
{\rm scheme} & $\,\alpha_s(M_Z)$\, & $\,\alpha_0(2~{\rm GeV})$\, & $\chi^2$/d.o.f.\,
\\[0.2em]\hline
\phantom{x} & &
\\[-1em]
{\rm log-R} & $0.1160^{+0.0050}_{-0.0051}$ & $0.565^{+0.109}_{-0.085}$\, & $1.44$\,
\end{tabular}
\end{center}

whose central value is consistent with~\cite{davisonwebber, Bethke:2011tr} but the uncertainties are larger as expected.

\section{Conclusions}
\label{sec:Conclusions}
In the present paper we performed a new fit of the strong coupling $\alpha_s$ using the event shape observable thrust. For the theoretical predictions we use resummed NNLL results matched to fixed
order NNLO calculations. For the hadronisation corrections we use an analytical dispersive model, extending the leading renormalon subtraction to ${\cal O}(\alpha_s^3)$. Furthermore we include finite
bottom quark mass effects to
NLO accuracy. Together with the perturbative coupling $\alpha_s$, a simultaneous fit of the non-perturbative mean effective coupling $\alpha_0$ was also performed.

For the $\alpha_s$ determination we used experimental data in the energy range $14<Q<209$ GeV which have been corrected for QED radiation effects from ALEPH, L3 and the TASSO experiments. The
theoretical uncertainties are determined as the envelope of all variations obtained by performing a parameter scan in which we vary the renormalization scale $x_{\mu}$, the resummation scale $x_{L}$,
the modified-logarithms parameter $p$ and the
Milan factor ${\cal M}$. Furthermore we consider bottom-quark mass corrections for three different $b$-quark masses $m_b=4.0$ GeV, $m_b=4.5$ GeV and $m_b=5.0$ GeV. The scans are performed using both
the
R matching scheme, leading to the following values
\begin{align}
\alpha_s(M_Z)=&0.1131^{+0.0028}_{-0.0022},\quad\alpha_0(2~{\rm GeV})=&0.538^{+0.102}_{-0.047}\,,\nonumber
\end{align}
and the log-R matching scheme which leads to
\begin{align}
\alpha_s(M_Z)=&0.1137^{+0.0034}_{-0.0027},\quad\alpha_0(2~{\rm GeV})=&0.524^{+0.096}_{-0.044}\,.\nonumber
\end{align}
In both cases the quoted central value refers to the default setting of the theory parameters.
The obtained values are compatible with recent determinations of $\alpha_s$ from thrust using effective field theory results~\cite{Abbate:2010xh,Abbate:2012jh}, but the uncertainty is larger. 
Compared to previous determinations of the strong coupling, where Monte Carlo hadronisation corrections were used, the present determination leads to smaller values for $\alpha_s$. 
This can be mainly explained by observing that Monte Carlo models are currently tuned on experimental data using a showered (to leading logarithmic accuracy) leading order perturbative prediction. 
The effect of the NNLL resummation on the determination of $\alpha_s$ has been discussed in Section~\ref{sec:Comparison}, and it shows that higher order logarithmic terms have a sizeable
impact on its final value. This suggests that Monte Carlo hadronisation should not be used together with higher order resummed predictions since this would misestimate the real hadronisation
corrections. This issue has been already discussed in~\cite{Dissertori:2009ik,jaquier}. 
Our final value for $\alpha_s$ is also smaller than the current world average~\cite{Bethke:2012zza}, whose determination involves many results from both lattice computations and $\tau$-decay, which
lead to higher values of $\alpha_s$.
In view of recent developments in extending the resummation beyond NLL accuracy for event-shape variables other than thrust, it would be interesting to repeat a similar analysis for the full set of
six event-shape observables to obtain a more robust fit of the strong coupling constant.

\section{Acknowledgements}
We are grateful to G\"unther Dissertori, Gavin Salam and Hasko Stenzel for valuable discussions.
We thank Carlo Oleari for providing us with an up-to-date version of the code {\tt Zbb4}, 
and Andreas Papaefstathiou for helpful discussions about {\tt Herwig++}.
G.L. would like to thank the Institute for Theoretical Physics, University of Zurich for the warm hospitality while part of this work was carried
out.\\
G.L. was supported by the British Science and Technology Facilities Council (STFC) and by the Alexander von
Humboldt Foundation, in the framework of the Sofja Kovaleskaja Award Project
``Advanced Mathematical Methods for Particle Physics'', endowed by the German
Federal Ministry of Education and Research. T.G. and P.F.M. were supported by the Swiss National Science Foundation (SNF) under grant 200020-138206 and the European Commission through the
LHCPhenoNet network under contract PITN-GA-2010-264564.

\appendix

\section{Explicit resummation formulae}
\label{sec:resummformul}
In the present section we report the full resummation formulae used in the main text.
The QCD $\beta$-function is defined by the renormalisation group equation for the QCD coupling constant
\begin{align}
\label{rgealpha}
\frac{d\alpha_{s}(\mu)}{d\ln\mu^{2}}=-\alpha_{s}(\mu)\bigg(\frac{\alpha_{s}(\mu)}{\pi}\beta_{0}+\frac{\alpha_{s}^{2}(\mu)}{\pi^{2}}\beta_{1}+\ldots \bigg),
\end{align}
where the first two coefficients read
\begin{align}
\beta_{0} =& \frac{11}{12}C_{A}-\frac{1}{3}T_{F}n_{F},\notag\\
\beta_{1} =& \frac{17}{24}C_{A}^{2}-\frac{5}{12}C_{A}T_{F}n_{F}-\frac{1}{4}C_{F}T_{F}n_{F}.
\end{align}
Following these conventions, the perturbative functions $h_i(\alpha_s L)$ have the following expressions~\cite{CTTW,Monni:2011gb}, as a function of
$\lambda=\frac{\alpha_{s}(\mu)}{\pi}\beta_{0}\log N$
\begin{align}
\label{f1}
h_1(\alpha_s L) &= -\frac{A^{(1)}}{\beta_{0}\lambda}[(1-2\lambda)\ln(1-2\lambda)\notag\\
&-2(1-\lambda)\ln(1-\lambda)]\,,
\end{align}
\begin{align}
h_2(\alpha_s L) &= -\frac{A^{(2)}}{\beta_{0}^{2}}[2\ln(1-\lambda)-\ln(1-2\lambda)]\notag\\
&+2\frac{B^{(1)}}{\beta_{0}}\ln(1-\lambda)
-\frac{A^{(1)}\beta_{1}}{\beta_{0}^{3}}[\ln(1-2\lambda)\notag\\
&+\frac{1}{2}\ln^{2}(1-2\lambda)-\ln(1-\lambda)(2+\ln(1-\lambda)]\notag\\
\label{f2}
&-2\frac{A^{(1)}\gamma_{E}}{\beta_{0}}\ln\frac{1-\lambda}{1-2\lambda}\,,
\end{align}
\begin{align}
h_3(\alpha_s L) &=\,\,\,-\frac{2B^{(2)}}{\beta_{0}^{2}}\frac{\lambda}{1-\lambda}-\frac{A^{(3)}}{\beta_{0}^{3}}\frac{\lambda^{2}}{(1-\lambda)(1-2\lambda)}\notag\\
&-\frac{2A^{(2)}\gamma_{E}}{\beta_{0}^{2}}\frac{\lambda}{(1-\lambda)(1-2\lambda)}\notag\\
&+\frac{A^{(2)}\beta_{1}}{\beta_{0}^{4}}\bigg[\frac{3\lambda^{2}}{
(1-\lambda)(1-2\lambda)}+\frac{\ln(1-2\lambda)}{1-2\lambda}\notag\\
&-2\frac{\ln(1-\lambda)}{1-\lambda}\bigg]-2 \frac{B^{(1)}}{\beta_{0}}\gamma_{E}\frac{\lambda}{1-\lambda}\notag\\
&+\frac{2B^{(1)}\beta_{1}}{\beta_{0}^{3}}\frac{\lambda+\ln(1-\lambda)}{1-\lambda}+\frac{A^{(1)}}{\beta_{0}}\frac{1}{(1-\lambda)(1-2\lambda)}\notag\\
&\times\bigg[
\frac{2\gamma_{E}\beta_{1}}{\beta_{0}^{2}}\big[\lambda+(1-\lambda)\ln(1-2\lambda)\notag\\
&-(1-2\lambda)\ln(1-\lambda)\big]-\gamma_{E}^{2}\lambda(3-2\lambda)\notag\\
&+\frac{\beta_{2}}{\beta_{0}^{3}}\big[-\lambda^{2}
+(1-\lambda)(1-2\lambda)\ln\frac{(1-\lambda)^2}{1-2\lambda}\big]\bigg]\notag\\
&-\frac{A^{(1)}\beta_{1}^{2}}{\beta_{0}^{5}}\bigg[\frac{1-\lambda}{2(1-\lambda)(1-2\lambda)}\ln(1-2\lambda)\notag\\
&\times\big[4\lambda+\ln(1-2\lambda)\big]-\frac{1}{(1-\lambda)(1-2\lambda)}\notag\\
\label{f3}
&\times\big[\lambda^{2}-(1-2\lambda)\ln(1-\lambda)(2\lambda+\ln(1-\lambda))\big]\bigg]\,.
\end{align}
The resummation coefficients read
\begin{subequations}
\begin{align}
A^{(1)} &= C_F, \qquad 
\frac{A^{(2)}}{A^{(1)}} =  \left(\frac{67}{36}-\frac{\pi^2}{12}\right)C_{A}-\frac{5}{18}n_{F},\\
\frac{A^{(3)}}{A^{(1)}} &= n_{F}^2\left(\frac{25}{324}-\frac{\pi^2}{216}\right)+C_{A}n_{F}\left(-\frac{2051}{1296}+\frac{7}{72}\pi^2\right)\nonumber\\
&+C_{A}^2\bigg(\frac{15503}{2592}-\frac{389}{864}\pi^2+\frac{11}{720}\pi^4-\frac{11}{4}\zeta_{3}\bigg)\nonumber\\
&+C_{F}n_{F}\left(-\frac{55}{96}+\frac{\zeta_{3}}{2}\right)\,,
\end{align}
\begin{align}
B^{(1)} &= -\frac{3}{4}C_F,\\
B^{(2)} &= C_{F}n_{F} \left(\frac{1}{48}+\frac{\pi^{2}}{36}\right)+C_{F}^{2}\left(-\frac{3}{32}+\frac{\pi^{2}}{8}-\frac{3 \zeta_{3}}{2}\right)\notag\\
&+C_{A}C_{F}\left(-\frac{17}{96}-\frac{11\pi^{2}}{72}+\frac{3\zeta_{3}}{4}\right)\,.
\end{align}
\end{subequations}
Additional contributions to the functions $g_i(\alpha_sL)$~(\ref{eq:csigma2}) arise from the function $H(1,\alpha_{s}(Q))$ which accounts for hard virtual corrections as well as the hard
collinear $\tilde{J}\left(1,\alpha_{s}(\sqrt{\frac{N_{0}}{N}}Q)\right)$ and soft $\tilde{S}\left(1,\alpha_{s}(\frac{N_{0}Q}{N})\right)$ constant functions. The latter functions contribute to the
perturbative logarithmic structure since their strong couplings are evaluated at the collinear ($\sqrt{\frac{N_{0}}{N}}Q$) and soft ($\frac{N_{0}Q}{N}$) scales, respectively.  Their perturbative
expansions can be found for example in the appendix of~\cite{Becher:2008cf,Monni:2011gb}.
  
\newpage

\end{document}